\documentclass[aps,superscriptaddress,amsfonts,amssymb,amsmath,amsmath,nofootinbib,twocolumn]{revtex4}

\usepackage[utf8]{inputenc}
 \usepackage[T2A]{fontenc}
 \usepackage[english]{babel}
\usepackage{graphicx,color}
\definecolor{darkblue}{rgb}{0,0,0.7}
\definecolor{darkred}{rgb}{0.7,0,0}
\usepackage[unicode, colorlinks, citecolor=darkblue, linkcolor=darkred, urlcolor=blue]{hyperref}

\usepackage{euler}

\allowdisplaybreaks

\begin{document}

\title{Polarization loss in reflecting coating}

\author{S. P. Vyatchanin}

\affiliation{Faculty of Physics, Moscow State University, Moscow, 119991 Russia,\\
             Quantum Technology Centre, Moscow State University, Moscow, 119991 Russia}

\date{January 29, 2020}

\begin{abstract}
 In laser gravitational waves detectors optical loss restricts sensitivity. We discuss polarization scattering as one more possible mechanism of optical losses. Circulated inside interferometer light is polarized and after reflection its plane of polarization can turn a little due to reflecting coating of mirror can have slightly different refraction index along axes $x,\, y$ in plane of mirror surface (optical anisotropy).  This anisotropy can be produced during manufacture of coating (elasto-optic effect). This orthogonal polarized light, enhanced in cavity, produces polarization optical loss. Polarization map of mirrors is very  important and we propose to measure it. Polarization loss can be important in different precision optical experiments based on usage of polarized light, for example, in quantum speed meter.
\end{abstract}

\maketitle

\section{Introduction}

The advanced LIGO detectors directly detected  gravitational waves (GW) \cite{GW_2016, GW_2016b, GW_2017, GW_2017b} and  opened the era of gravitational wave astronomy. This detection became possible due to technology breakthroughs allowing to measure very small displacements of $\sim 10^{-18}$\,m of test masses \cite{aLIGO2013, aLIGO2014, aLIGO2015, aLIGO2015b, aVirgo2015, grote2010} separated by distance 4 km. The main limitation for the detector sensitivity in  most sensitive frequency band from $50$~Hz to $2000$~Hz \cite{99a1BrLeVy, 99a1BrGoVy, 00a1BrGoVy, HarryBook2012} is produced by the thermal noise of the test masses. Currently, a main source for thermal noise is Brownian structural noise \cite{nawrodt2011} in mirror coatings. Random stresses, thermally induced  in the substrates and the coatings  of the test masses, originate random deformations of their surfaces, which are detected at the interferometer output as thermal noise.

Optical losses of mirror --- one more noise, restricting sensitivity of GW detectors. Optical losses prevents achievement of Standard Quantum Limit \cite{Braginsky68} and deteriorates optical squeezing \cite{02a1KiLeMaThVy, MiaoPRX2019, EvansPRD2013}. Optical losses are mainly originated by scattering on roughness of mirror surface and imperfections of layers in interferometric coatings. In Advanced LIGO round trip losses in cavities in arms are about 100 ppm, in GW detectors of third generation these losses are planed to be lees, about 60 ppm \cite{3dGen2019}.  

In this article we discuss one more possible mechanism of optical losses --- polarization scattering produced by elasto-optic effect in reflecting coating. (Note, polarization scattering in substrates is also important, in particular it was analyzed in \cite{KrugerArXiv2015}, however, it is not a subject of this paper.)
 In GW detectors the light, circulated inside interferometer, is polarized. In ideal case after reflection from mirror polarization of reflected light is conserved. However, reflecting coating can have slightly different refraction index along axes $x,\, y$ in plane of mirror surface. In other words elasto-optic effect can produces optical anisotropy of coating.  This anisotropy can be produced during manufacture of coating. Then polarization of reflected light will be slightly turned. This small orthogonal polarized light, enhanced in cavity, produces polarization optical losses.

\section{Rough estimates of elasto-optic effect}

The interferometric coating is evaporated on end (or input) mirrors. The temperature of coating during evaporation is about $T_e\simeq 600 \dots 700\, \text{C}^\circ \simeq 900 \dots 1000\, \text{K}^\circ$, so at room temperature $T_0\simeq 300$\,K the coating will be stressed. Its deformation $\Delta \ell/\ell$ can be estimated as
\begin{align}
 \frac{\Delta \ell}{\ell} \simeq \big(\alpha_c - \alpha_s\big)\big[T_e-T_0\big] \simeq 6\cdot 10^{-3}
\end{align}
where $\alpha_c,\ \alpha_s$ are thermal expansion coefficients of coating and substrate, here we put $\alpha_c-\alpha_s\simeq 10^{-5}\, \text{K}^{-1}$. 


In ideal case the coating film is stressed uniformly, i.e. each element of coating is stressed equally in all direction  and deformation $\Delta \ell/\ell$ is the same in all directions. It means that in this ideal case refraction index is the same for any deformations, i.e. 
\begin{align}
 n_- &\equiv n_x -n_y=0,\quad n_+\equiv n_x+n_y =\text{const},
\end{align}
where $n_x,\, n_y$ are refraction indexes in coating for light polarized along axes $x,\, y$ on mirror surface. The axes $x,\ y$ are chosen so that $n_x$ ($n_y$) has maximal (minimal) value.

However, non-uniform heating during evaporation can create anisotropy in stressed coating. Then at room temperature the coating should consists of pieces in which  deformation has different value and direction. One can estimate deformation in such piece through variation of temperature $\Delta T\simeq 100\, \text{K}$ of coating during manufacture
\begin{align}
 \frac{\Delta \ell_a}{\ell} \simeq \big(\alpha_c - \alpha_s\big)\, \Delta T \simeq 1\cdot 10^{-3}
\end{align}


Birefringence induces phase shift $\phi$ between ordinary and extraordinary waves:
\begin{align}
\label{phi}
 \phi \simeq 2n_- kz_\text{eff}\simeq 5\cdot 10^{-3}.
\end{align}
Here we assume that $kz_\text{eff}\simeq \pi$ (it corresponds to two quarter wavelength layers of coating), $\phi^2$ gives the estimate of relative power redistributed from the main to orthogonally polarized mode.

This estimate is rather large. Indeed, estimate \eqref{phi} shows that after reflection about $\phi^2\simeq 25$ ppm of incident light will re-emitted in orthogonal polarization. 

We can not point the possible size of isotropically stressed domains of coating --- it is a subject of further experiment.  We have two possibilities: large and small domains.

For large domains the sizes of anisotropic stressed domains are comparable with radius of light beam spot. In this case the main part of orthogonally polarized light will be emitted into main optical mode cavity. Hence, it will be resonantly enhanced and in light reflected from FP cavity the part of orthogonally polarized light will be enlarged. Indeed, its amplitude $B_\bot$ is equal to:
  \begin{align}
   B_\bot = \frac{2}{T}\cdot B\phi\,.
  \end{align}
  Here $T$ is {\em power} transmittance of input mirror of FP cavity (we assume perfect reflectivity of end mirror), $B$ is an amplitude $B$  of incident light. We see that $B_\bot$ is larger than $ \phi B$ (estimate based on \eqref{phi}) by resonance factor $2/T$. Then relative losses into orthogonally polarised light can be estimated as
  \begin{align}
   \mathcal L_{reson1} &\simeq \left(\frac{2}{T}\cdot \phi\right)^2 \simeq 0.05\ (!)
  \end{align}
   Here we use estimate \eqref{phi} and put $T =4\cdot 10^{-2}$.  
   
   Obviously, possibility  of large domains existence  is very low. 
  
   For another case of small domains the sizes of anisotropic stressed domains are much less than radius of light beam spot. In this case the main part of polarised light will be emitted quasi-isotropically into space. It means that estimate of ``polarised'' losses in this case can be estimated as $\mathcal L_{isotropic}\simeq\phi^2$. However its small part $\eta$ will be re-emitted in main mode and will be resonantly enhanced. Hence in this case the ``polarized'' losses have two part: quasi-isotropic $\mathcal L_{isotropic}$ and resonantly enhanced $\mathcal L_{reson2}$:
   \begin{align}
    \mathcal L_{isotropic} &\simeq \phi^2,\quad \mathcal L_{reson2} \simeq \left(\frac{2\eta}{T}\cdot \phi\right)^2  
   \end{align}
  The coefficient $\eta$ can be calculated from map $n_-(\vec r)$ of difference refraction index $n_-$. Obviously, fraction $\eta/T$ can be larger than unity.  

The case of small domains looks more  probable. Then the key problem is to measure birefringence map of coating of each mirror.

\begin{figure}
	\includegraphics[width=0.3\textwidth]{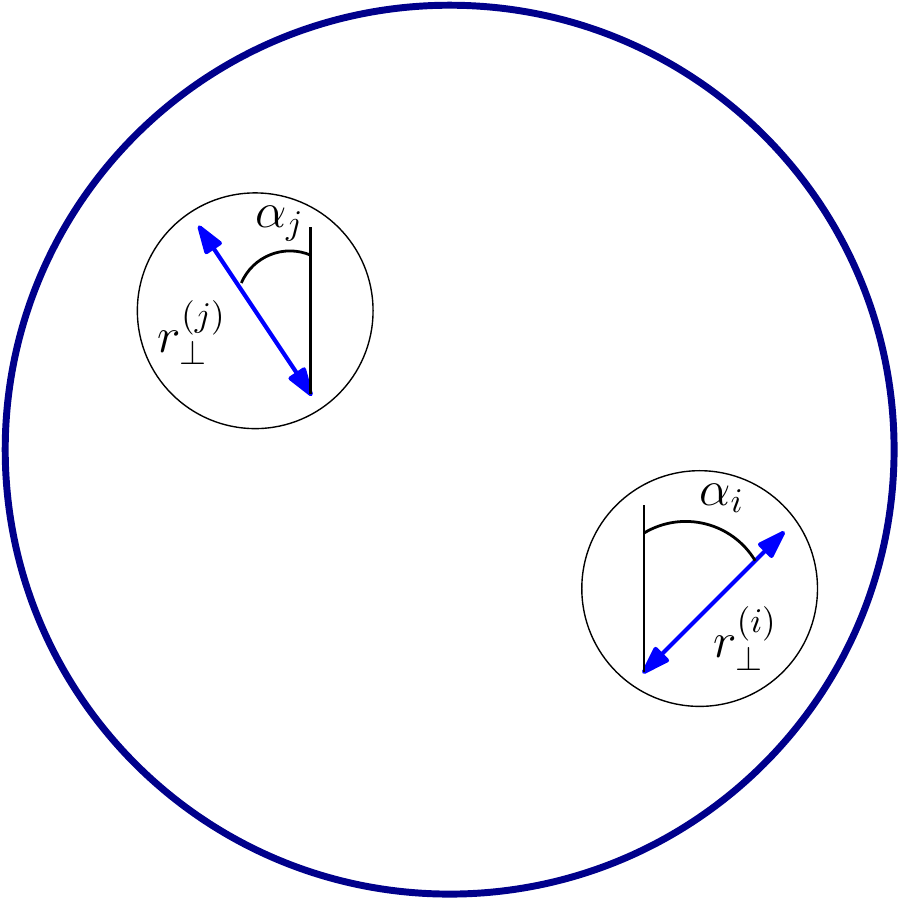}
	\caption{The physical volumes of mirror coating have different values and directions of stresses. The incident is polarized vertically.  We characterise each volume by difference refraction indexes $n_-^{(j)}$ and  $\alpha_j$.}\label{domains}
\end{figure}

\section{Model}

The diameter of mirrors in Advanced LIGO GW detectors are about 40 cm, the diameters of mirrors in GW detectors of third generation will be larger. We assume that small physical pieces of mirror coating have different values and directions of stresses, producing small birefringence as shown on Fig.~\ref{domains}.  This birefringence  can be detected by reflection of  polarized light focused on small square $\delta S$ on coating. The polarization of reflected light will slightly differ from polarization of incident light having small component $\delta A_\bot$ of orthogonal polarization.  In general case its amplitude is equal to
\begin{align}
  \delta A_\bot^{(j)} = R_\bot^{(j)} \sin 2\alpha_j\, A
\end{align}
Here $R_\bot^{(j)}$ ``orthogonal'' amplitude reflection coefficient of $j$-th piece (with square $\delta S$) on coating surface, $\alpha_j$ is angle between direction of incident light polarization  and stress direction (birefringence axis) --- see also Fig.~\ref{domains}. Obviously, small orthogonal polarization reaches maximum  at optimal angle $\alpha_j= \pi/4$.

In general case one can assume that birefringence parameters of coating are function of coordinates $\vec r$ on mirror surface: $R_\bot(\vec r),\ \alpha(\vec r)$. They are different for each mirror and can be considered as random variable


Let incident light is vertically polarized and has Gaussian distribution in cross section (main mode), its  electrical field $E (\vec r)$ is equal to:
\begin{align}
 E (\vec r)&= A \, \Psi (\vec r),\\
 \Psi(\vec r) &=\frac{e^{-r^2/2r_0^2}}{\sqrt{\pi r_0^2}},\quad \int \Psi^2(\vec r)\, d\vec r =1,
\end{align}
where $A$ is an amplitude of electrical field , $r_0$ is effective radius of light beam on mirror, integration is taken over mirror surface. Then amplitude of reflected orthogonal polarized light, emitted into main mode of cavity, is equal to
\begin{align}
 E_\bot (\vec r)&= A \int \Psi^2 (\vec r)\, R_\bot(\vec r) \sin\big[2\alpha(\vec r)\big]\, d \vec r\,. 
\end{align}
The average value $\langle E_\bot\rangle =0$ ($R_\bot(\vec r),\ \alpha(\vec r)$ are random value) and it is  variation $\Delta e_\bot^2$ that should characterize relative orthogonal polarization component:
\begin{align}
 \Delta e_\bot^2 &=\frac{\langle E^2_\bot(\vec r)\rangle}{A^2} =\\
	 & = \left\langle \left[\int \Psi^2 (\vec r)\, R_\bot(\vec r) \sin\big[2\alpha(\vec r)\big]\, d \vec r\right]^2
	 \right\rangle\,.
\end{align}
Hence, we can estimate value of orthogonal polarized amplitude
\begin{align}
 \label{Ebot}
 E_{\bot 00} &\simeq A\sqrt{ \Delta e_\bot^2}\,. 
\end{align}
Underline that mirror is a part of Fabry-Perot cavity, hence, the light emitted into main mode wave with amplitude $E_{\bot 00}$  will be resonantly enhanced. It is easy to show that in resonance case amplitudes of wave inside cavity  $E_{\bot 00}^\text{in}$ and outside cavity $E_{\bot 00}^\text{out}$ are equal to:
\begin{align}
 E_{\bot 00}^\text{in} &= \frac{E_{\bot 00}}{1-\sqrt R}\simeq \frac{2E_{\bot 00}}{T},\quad
	 E_{\bot 00}^\text{out} = \sqrt T\, E_{\bot 00}^\text{in}
\end{align}
Here $R,\ T$ are power reflectivity and transmittance of input mirror, $T\ll 1$. End mirror is assumed to be perfectly reflective. Pay attention that $A$  in \eqref{Ebot} is mean amplitude inside cavity, amplitude of wave outside cavity is
\begin{align}
 A^\text{out} &= \sqrt T\, A.
\end{align}
and ratio of orthogonal polarized amplitude to mean amplitude outside cavity is equal to
\begin{align}
 \frac{E_{\bot 00}^\text{out}}{A^\text{out}} & = \frac{2}{T}\, \sqrt{ \Delta e_\bot^2}
	 = \frac{\sqrt{ \Delta e_\bot^2}}{\gamma\tau},\quad \gamma =\frac{T}{2\tau},
\end{align}
where $\gamma$ is cavity relaxation time, $\tau =L/c$ is round trip time. Even for $\Delta e_\bot^2\simeq 1$~ppm and $T=4\cdot 10^{-2}$ we estimate 
\begin{align}
 \left[\frac{E_{\bot 00}^\text{out}}{A^\text{out}}\right]^2 \simeq 2.5\cdot 10^{-3} = 2.5\, \text{pm}
\end{align}

We see that Fabry-Perot cavity considerably  enhances polarization loss (and noise) in main mode.

\section*{Discussion and Conclusion}

In experiments with polarized light the appearance of light of orthogonal polarization means additional loss which can be called as polarization loss. For account of this polarization loss measurement of the polarization map $R_\bot(\vec r),\ \alpha (\vec r)$ is very important.

\begin{figure}
	\includegraphics[width=0.25\textwidth]{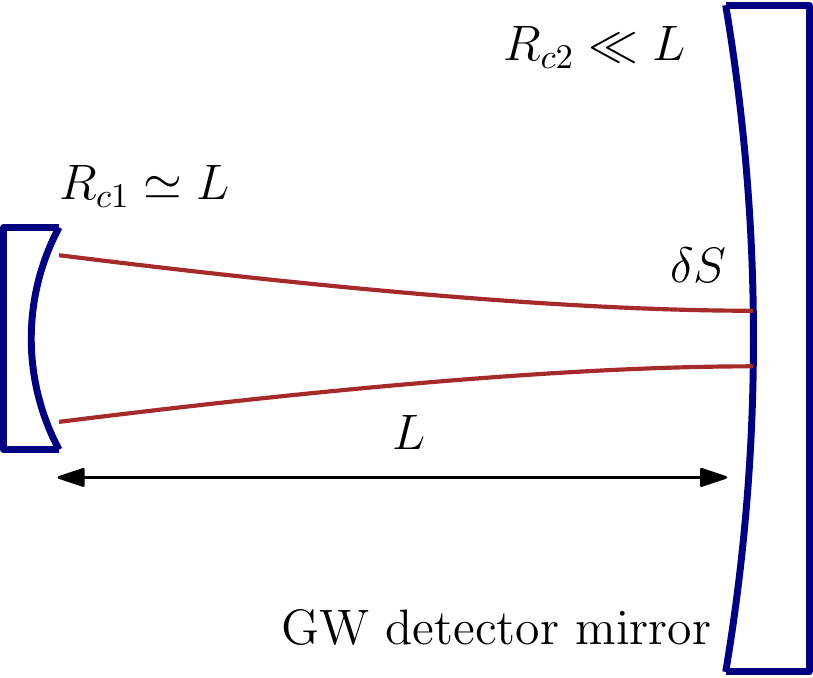}
	\caption{The possible scheme for measurement of polarization map of GW detector mirror. Orthogonally polarized light, appeared after reflection, is resonantly enhanced in cavity.}\label{map}
\end{figure}

Note that in accordance of Fluctuation-Dissipation Theorem any losses produces additional noise, and hence, decreases experimental  sensitivity.

Polarization loss (and noise) can be important for laser interferometric GW detectors because it means additional noise produced by light of orthogonal polarization in dark port. Another example, in order to surpass Standard Quantum Limit in GW detectors there were proposed the  quantum speed meter based on polarization schemes \cite{12a1DaKh, 19a1DaKhMi}, and it is polarization noise which will restrict sensitivity of this speed meter. 

For mirrors of laser interferometric GW detectors the roughness of surfaces are very important and measurement of  roughness maps are performed now. We would like to pay attention that polarization map $R_\bot(\vec r),\ \alpha (\vec r)$ of these mirrors has not smaller importance and should be also performed.

Polarization map measurement is not a simple task and in order to increase accuracy through resonant enhancement we propose to use cavity scheme shown on Fig.~\ref{map}. Cavity is assembled by large mirror of GW detector and small additional mirror which curvature radius $R_{c1}$ is about distance $L$ of cavity. Mirrors of GW detector have large curvature radius $R_{c2}$ about several kilometers, so $R_{c2}\gg L$. Hence, it is approximately hemispherical cavity which waist is practically placed on the surface of GW detector mirror and can have size comparable with wave length $\lambda$. Rotating polarization of pump one can measure values $R_\bot$ and $\alpha$ of spot (placed in waist of beam). Scanning light beam over surface of GW detector mirror one can measure polarization map. 

Obviously, pump laser must have stable polarization. Also, small additional mirror has to have no polarization scattering or, at least, it should be known with high accuracy.  

Note that polarization scattering in mirror substrate (silicon \cite{KrugerArXiv2015} in Einstein Telescope or Cosmic Explorer or sapphire in KAGRA) can  also be important reason of additional losses and it should be carefully investigated.

Summing up, polarization loss (and noise) can be important in precision optical experiment and it should be investigated.

\acknowledgments
 Author acknowledges partial support from the Russian Foundation for Basic Research (Grant No. 19-29-11003) and from the TAPIR GIFT MSU Support of the California Institute of Technology. This document has LIGO number LIGO-P2000038.


\end{document}